\documentclass[aps,prd,twocolumn,nofootinbib,superscriptaddress,preprintnumbers,floatfix]{revtex4-2}
\usepackage{float,dcolumn}
\usepackage{amsmath,amssymb}
\usepackage{mathtools,slashed}
\usepackage{bm}
\usepackage{xcolor}
\usepackage[caption=false]{subfig}
\usepackage{verbatim}
\usepackage{array}
\usepackage{multirow}
\usepackage{booktabs}
\usepackage{url}
\usepackage{color}
\usepackage[T1]{fontenc}
\usepackage[latin9]{inputenc}
\usepackage{graphicx}
\usepackage{esint}
\usepackage[hypertexnames=false]{hyperref}
\usepackage{cleveref}
\usepackage{atbegshi}

\usepackage[normalem]{ulem}

\Crefname{equation}{Eq.}{Eqs.}
\Crefname{section}{Sec.}{Secs.}
\Crefname{figure}{Fig.}{Figs.}
\Crefname{appendix}{Appendix}{Appendices}

\newcommand{\cK}[0]{\mathcal K}

\newcommand{\cM}[0]{\mathcal M}

\hypersetup{ 
pdfnewwindow=true,
colorlinks=true,    
linkcolor=blue,
citecolor=blue,
filecolor=blue,
urlcolor=blue
} 

\begin{document}

\preprint{\vbox{\hbox{MIT-CTP/5845} }}
\title{QCD predictions for physical multimeson scattering amplitudes } 

\author{Sebastian M. Dawid}
\affiliation{Physics Department, University of Washington, Seattle, WA 98195-1560, USA}

\author{Zachary T. Draper}
\affiliation{Physics Department, University of Washington, Seattle, WA 98195-1560, USA}

\author{Andrew D. Hanlon}
\affiliation{Department of Physics, Kent State University, Kent, OH 44242, USA}

\author{Ben H\"{o}rz}
\affiliation{Intel Deutschland GmbH, Dornacher Str. 1, 85622 Feldkirchen, Germany}

\author{Colin Morningstar}
\affiliation{Department of Physics, Carnegie Mellon University, Pittsburgh, Pennsylvania 15213, USA}

\author{Fernando Romero-L\'{o}pez}
\affiliation{Albert Einstein Center, Institute for Theoretical Physics, University of Bern, 3012 Bern, Switzerland}
\affiliation{Center for Theoretical Physics, Massachusetts Institute of Technology, Cambridge, MA 02139, USA}

\author{Stephen R. Sharpe}
\affiliation{Physics Department, University of Washington, Seattle, WA 98195-1560, USA}

\author{Sarah Skinner}
\affiliation{Department of Physics, Carnegie Mellon University, Pittsburgh, Pennsylvania 15213, USA}

\date{\today}

\begin{abstract}
We use lattice QCD calculations of the finite-volume spectra of systems of two and three mesons to determine, for the first time, three-particle scattering amplitudes with physical quark masses. Our results are for combinations of $\pi^+$ and $K^+$, at a lattice spacing $a=0.063\;$fm, and in the isospin-symmetric limit. We also obtain accurate results for maximal-isospin two-meson amplitudes, with those for $\pi^+ K^+$ and $2K^+$ being the first determinations at the physical point.
Dense lattice spectra are obtained using the stochastic Laplacian-Heaviside method, and
the analysis leading to scattering amplitudes is done using the relativistic finite-volume formalism. Results are compared to chiral perturbation theory and to phenomenological fits to experimental data, finding good agreement.
\end{abstract}

\maketitle

{\it Introduction.}---Predicting the properties and interactions of hadrons directly from Quantum Chromodynamics (QCD) is a formidable challenge with far-reaching implications---from the emergence of nuclear forces and the existence of exotic hadrons to precision tests of the Standard Model through hadronic decays. Although quark models~\cite{Gell-Mann:1964ewy} and effective field theory approaches~\cite{Meissner:2025qdg,Mai:2025wjb} have provided significant insights, first-principles lattice QCD (LQCD) calculations have matured to the point that many
questions about hadron dynamics
can be addressed quantitatively~\cite{Briceno:2017max,Hansen:2019nir,Rusetsky:2019gyk,Mai:2021lwb,Romero-Lopez:2022usb,Bulava:2022ovd,Mai:2022eur}.

Existing LQCD calculations of multihadron processes have predominantly focused on two-particle decay channels and often used quark masses heavier than their physical values. By reducing the quark masses and incorporating decay modes involving more than two hadrons, it is possible to explore well-known three-pion resonances such as the $\omega(782)$ and $a_1(1260)$, exotic tetraquarks such as $T_{\rm cc}\to DD\pi$, the enigmatic Roper resonance, $N(1440) \to N\pi+N\pi\pi$, and even three-nucleon forces. The successful development of three-body tools for LQCD calculations~\cite{Detmold:2008gh,Beane:2007qr,Briceno:2012rv,Polejaeva:2012ut,Hansen:2014eka,Hansen:2015zga,Briceno:2017tce,Hammer:2017uqm,Hammer:2017kms,Mai:2017bge,Briceno:2018aml,Briceno:2018mlh,Pang:2019dfe,Jackura:2019bmu,Blanton:2019igq,Briceno:2019muc,
Romero-Lopez:2019qrt,Pang:2020pkl,Blanton:2020gha,Blanton:2020jnm,Romero-Lopez:2020rdq,Blanton:2020gmf,Muller:2020vtt,Blanton:2021mih,Muller:2021uur,Muller:2022oyw,Pang:2023jri,Bubna:2023oxo,Briceno:2024txg,Xiao:2024dyw,Hansen:2024ffk,Draper:2023xvu,Feng:2024wyg,Jackura:2023qtp,Briceno:2024ehy} has transformed these aspirations into a practical reality~\cite{Beane:2007es,Detmold:2008fn,Detmold:2008yn,Detmold:2011kw,Mai:2018djl,Horz:2019rrn,Blanton:2019vdk,Culver:2019vvu,Mai:2019fba,Fischer:2020jzp,Hansen:2020otl,Alexandru:2020xqf,Brett:2021wyd,Blanton:2021llb,NPLQCD:2020ozd,Mai:2021nul,Garofalo:2022pux,Yan:2024gwp,Draper:2023boj}.

Three-body calculations from LQCD have addressed increasingly complicated systems over the last decade.
We highlight three milestones from this active field.
First, the determination and analysis of a large number of $2\pi^+$ and $3\pi^+$ levels at $M_\pi\approx 200\;$MeV, showing the capabilities of LQCD techniques~\cite{Horz:2019rrn,Blanton:2019vdk,Mai:2019fba}.
Second, the calculation by the HadSpec collaboration of the $3\pi^+\to 3\pi^+$ scattering amplitude with $J^P=0^-$ at $M_\pi=391\;$MeV~\cite{Hansen:2020otl},
providing the first end-to-end application of the three-particle formalism. 
Third, the first constraints on the properties of three-particle resonances, 
specifically the width and mass of the $\omega(782)$~\cite{Yan:2024gwp}.
In this Letter, we present the results of a further benchmark calculation, namely that of three-meson scattering amplitudes at {\em physical quark masses}.
Furthermore, in addition to the dominant \(J^P=0^-\) amplitudes, we also determine for the first time those with \(J^P=1^+\) and \(2^-\).

Although our ultimate goal are three-particle resonances, we first consider nonresonant systems in order to work at physical quark masses, and examine all three-particle combinations of $\pi^+$ and $K^+$. The absence of resonances in these maximal-isospin systems simplifies the LQCD calculations (no quark annihilation contractions) and the finite-volume analysis (finite-volume spectrum close to the non-interacting case), and enables comparisons with chiral perturbation theory (ChPT).

A byproduct of three-body calculations is that they constrain the two-body amplitudes contributing via pairwise rescattering processes. Consequently, we accurately determine the two-particle amplitudes in maximal isospin channels---\(2\pi^+\), \(\pi^+ K^+\), and \(2K^+\)---the latter two computed for the first time with physical quark masses. When possible, we compare these two-meson amplitudes with phenomenological analyses of experimental data.

This Letter is released in parallel with a long article with all the details of the calculations~\cite{Dawid:2025doq}. Here we present the highlights and an outline of the methods.

{\it Scattering amplitudes from lattice QCD.}---LQCD calculations are carried out in Euclidean spacetime and in finite volumes. This complicates the extraction of scattering amplitudes, which are real-time infinite-volume observables. The finite-volume formalism provides the necessary connection, converting one set of physical quantities---finite volume energy levels---into another---the desired scattering amplitudes~\cite{Luscher:1986pf}.

In a finite volume with periodic boundary conditions, stationary energies deviate from the non-interacting case due to multihadron interactions. Intuitively, particles in a large but finite box are mostly separate, but undergo pairwise and triplet scatterings that shift the energy away from its free value. Although all partial waves contribute to this shift,  higher waves are suppressed in a threshold expansion, leading to a solvable inverse problem.

Energies are extracted from Euclidean correlation functions using a large set of operators with the system's quantum numbers. The resulting matrix of correlation functions is analyzed as a generalized eigenvalue problem~\cite{Luscher:1990ck}. Energies are obtained from the Euclidean-time dependence of the generalized eigenvalues. 

The first step in the finite-volume formalism employs two- and three-particle quantization conditions (QC2 and QC3). They describe all power-law $1/L^n$ finite volume effects, and are valid only up to corrections suppressed by $\exp(-M_\pi L)$. They are expressed as
\begin{align}
\det_{\Phi_n} \Bigl[ F_n\bigl( E^*_n, \bm{P}, L \bigr)^{-1} + \mathcal{K}_n(E_n^*) \Bigr] = 0\,,
\label{eq:QCN}
\end{align}
with \(n=2\) (\(n=3\)) corresponding to QC2 (QC3). These QCs depend on the total center of mass (c.m.) energy \(E^*_n\) of the \(n\)-body system, the total momentum \(\bm{P}\), and the box size \(L\). Each solution to the QCs corresponds to a finite-volume energy level. In the QC2, $\cK_2$ is the two-particle K matrix, an infinite-volume quantity that is related to the two-particle scattering amplitude, $\cM_2$. $F_2$ is a known finite-volume geometric function.
In the QC3, $\cK_{\rm 3}$ is a three-particle K matrix describing short-range three-body interactions, and $F_3$ is a known matrix that incorporates kinematic effects and two-particle interactions via $\cK_2$. The determinant is taken over indices, $\Phi_n$, that characterize the \(n\)-body state at fixed energy: two-body partial waves for the QC2, and  pair partial waves as well as the finite-volume momentum, $\bm p$, and flavor of the third, ``spectator'', particle for the QC3. The matrices in the QCs are finite, because we keep only two-body interactions up to a highest partial wave, $\ell \leq \ell_{\rm max}$, and, in the QC3, a cutoff in spectator momentum, $p \leq p_{\rm max}$. The latter condition defines a scheme for $\cK_3$, which is an unphysical intermediate object.

\begin{table}
   \centering
   \begin{tabular}{l c c c c c}
     \toprule
     & $(L/a)^3\times (L_t/a)$ & $M_\pi\;$[MeV] & $M_K\;$[MeV] & $M_\pi L$ & $N_{\rm cfg}$ \\
     \midrule
     E250 & $96^3 \times 192$ & 130 & 500 & 4.05 & 505 \\
     D200 & $64^3 \times 128$ & 200 & 480 & 4.20 & 2000 \\
     N200 & $48^3 \times 128$ & 280 & 460 & 4.42 & 1712 \\
     N203 & $48^3 \times 128$ & 340 & 440 & 5.41 & 771 \\
     \bottomrule
   \end{tabular}
   \caption{ Key parameters for ensembles used in this work. $N_{\rm cfg}$ is the number of gauge configurations. }
   \vspace{-0.1cm}
   \label{tab:ensembles}
\end{table}

In order to constrain $\cK_2$ and $\cK_3$, simple parametrizations of their kinematic dependence must be assumed. $\cK_2$ is parametrized as an expansion about threshold, using the standard effective-range expansion ($2K^+$) or a form incorporating the Adler zero ($2\pi^+$ and $\pi^+K^+$)~\cite{Adler:1964um}. These forms are truncated, and in practice involve $2-4$ parameters. Partial waves beyond $\ell>2$ are assumed to vanish. Similarly, $\cK_{\rm 3}$ is parametrized by a generalized threshold expansion, using a truncation consistent with that of $\cK_2$~\cite{Blanton:2019igq}, and also includes $2-4$ parameters. 

For the systems studied here, the QCs are valid below the first inelastic threshold, i.e. the c.m.~energy at which more than two (for the QC2) or three (for the QC3) particles can be on shell. For $2\pi^+$ and $3\pi^+$ the first inelastic threshold corresponds to adding two pions due to G parity. For all other systems with at least one kaon, the inelastic threshold is obtained by adding a pion. We restrict our fits to the regions of validity, although we find that the QCs work well for some range above.

The second step in the finite-volume formalism involves using the K matrices as input into integral equations whose solutions are the various three-particle amplitudes, $\cM_3$~\cite{Hansen:2015zga,Blanton:2021mih}. 
These equations produce amplitudes that automatically satisfy unitarity~\cite{Briceno:2019muc, Jackura:2022gib}.
They are solved as in Refs.~\cite{Jackura:2020bsk,Dawid:2023jrj,Dawid:2023kxu,Dawid:2024dgy}, with results discussed below.

\begin{figure*}[th!]
    \centering
    \includegraphics[width=\textwidth]{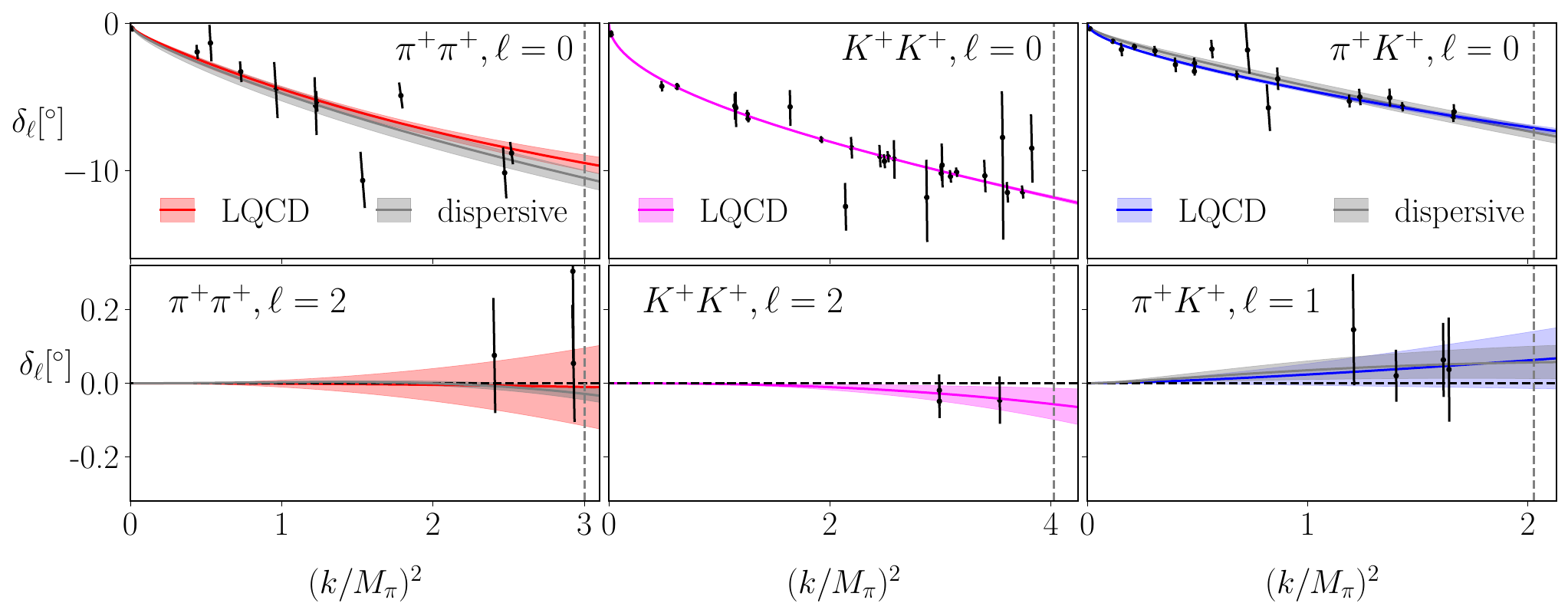} \vspace{-0.33cm}
    \caption{LQCD results for physical two-meson scattering phase shifts as a function of scattering momentum, $k$, in the lowest two partial waves for each system. Errors are statistical. Dispersive results~\cite{Garcia-Martin:2011iqs,Pelaez:2020gnd} are also shown. The vertical dashed lines indicate the inelastic threshold, above which the analysis breaks down. The ``data points'' are explained in the text.}
    \label{fig:2body}
\end{figure*}

{\it LQCD details.}---We use the $N_f=2+1$ CLS LQCD ensembles~\cite{Bruno:2014jqa} listed in \Cref{tab:ensembles}, with nonperturbatively $\mathcal{O}(a)$-improved Wilson fermions and the tree-level $\mathcal{O}(a^2)$-improved L\"uscher-Weisz gauge action. These have a fixed lattice spacing, $a \simeq 0.063\;$fm, isospin symmetry ($m_u=m_d\equiv m_\ell$), and lie on trajectory with constant $2m_\ell+m_s$.
The main results of this work use the E250 ensemble, which has approximately physical pion and kaon masses. We neglect the slight mistuning of the quark masses and treat them as physical. All ensembles have $M_\pi L>4$, leading to small exponentially-suppressed effects.

Energy levels are obtained for $2\pi^+$, $\pi^+K^+$, $2K^+$, $3 \pi^+$, $2\pi^+ K^+$, $2K^+\pi^+$, and $3 K^+$ channels, using several fitting approaches and two independent analyses. Correlated errors are obtained using the jackknife method. The levels are fit using the QC2 and QC3 to determine the K-matrix parameters.  Details of our most extensive fits are given in \Cref{tab:fits}, all of which have a $\chi^2$ per degree of freedom (DOF) lower than 2. 

\begin{table}
   \centering
   \begin{tabular}{l c c c}
     \toprule
     channels & $\#$ levels & $\#$ parameters & $\chi^2/{\rm DOF}$ \\
     \midrule
    $2\pi + 3\pi$ & $34+32$ & 6  & 1.14 \\
     $2\pi + \pi K + 2\pi K$ \ \ & $24+25+23$ & 9 & 1.23 \\
     $2K + \pi K + 2K\pi$ \ \ & $25+40+50$ & 10 & 1.95 \\
      $2K+3K$ & $40+53$ & 6 & 1.49 \\
     \bottomrule
   \end{tabular}
   \caption{Examples of fits. Particle charges are omitted. }
   \vspace{-0.1cm}
   \label{tab:fits}
\end{table}

{\it Two-meson phase shifts.}---Our most accurate results are for the two-particle phase shifts. These are obtained from combined correlated fits to two- and three-particle spectra,
leading to smaller errors than fits to the two-particle spectra alone. Both $2K^+$ and $\pi^+K^+$ have been explored directly at the physical point for the first time, while the $2\pi^+$ system was previously explored in Ref.~\cite{Fischer:2020jzp}.

In \Cref{fig:2body}, we show the resulting phase shifts for the lowest two partial waves. To illustrate the range of energy levels that contribute, we display the results that would be obtained were one to keep only the dominant partial wave contributing to each energy level.
In this approximation, there is a one-to-one relation between each energy and the phase shift. The curves are not fits to these points, but include much more information.

As can be seen, we obtain results with small errors for the $s$-wave phase shifts, while those for the higher waves are consistent with zero within small errors,
with some indication for a repulsive interaction in the $2K^+$ case.
For $2\pi^+$ and $\pi^+K^+$, we can compare our results to those obtained using dispersive fits to the experimental data~\cite{Garcia-Martin:2011iqs,Pelaez:2020gnd}.
We find results that are consistent within errors, and note that our errors are comparable in size. For the $2K^+$ channel, there is no ``experimental'' result available, so our result is a first-principles prediction. 

\begin{figure}[th!]
     \centering
    \includegraphics[width=0.25\textwidth]{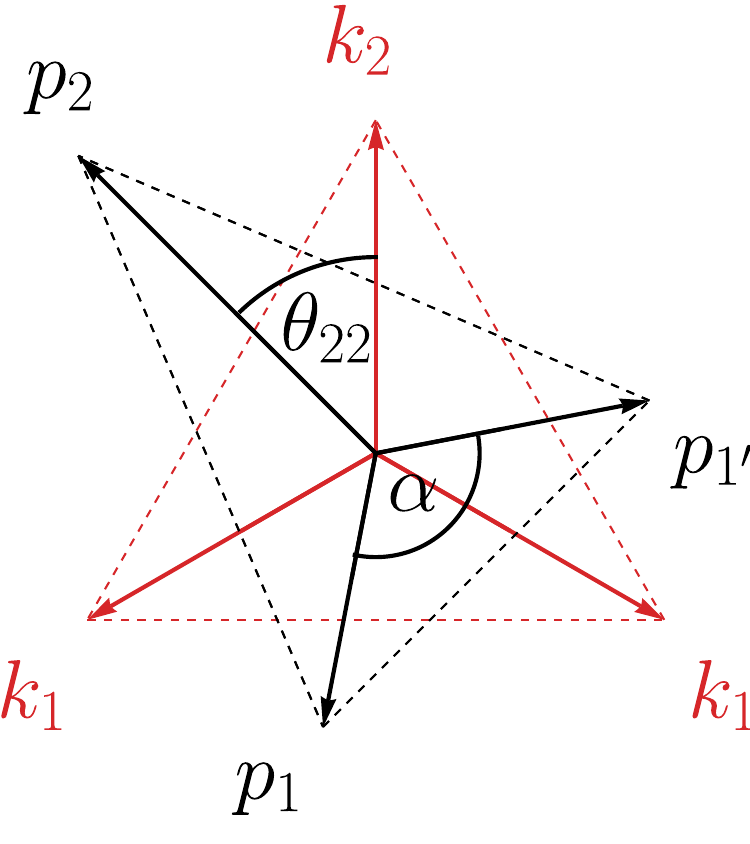}
    \vspace{-0.25cm}
    \caption{ Kinematic configuration used in this work. \label{fig:config}
    \vspace{-0.1cm}
    }
\end{figure}
\begin{figure}[th!]
     \centering
     \includegraphics[width=0.47\textwidth]{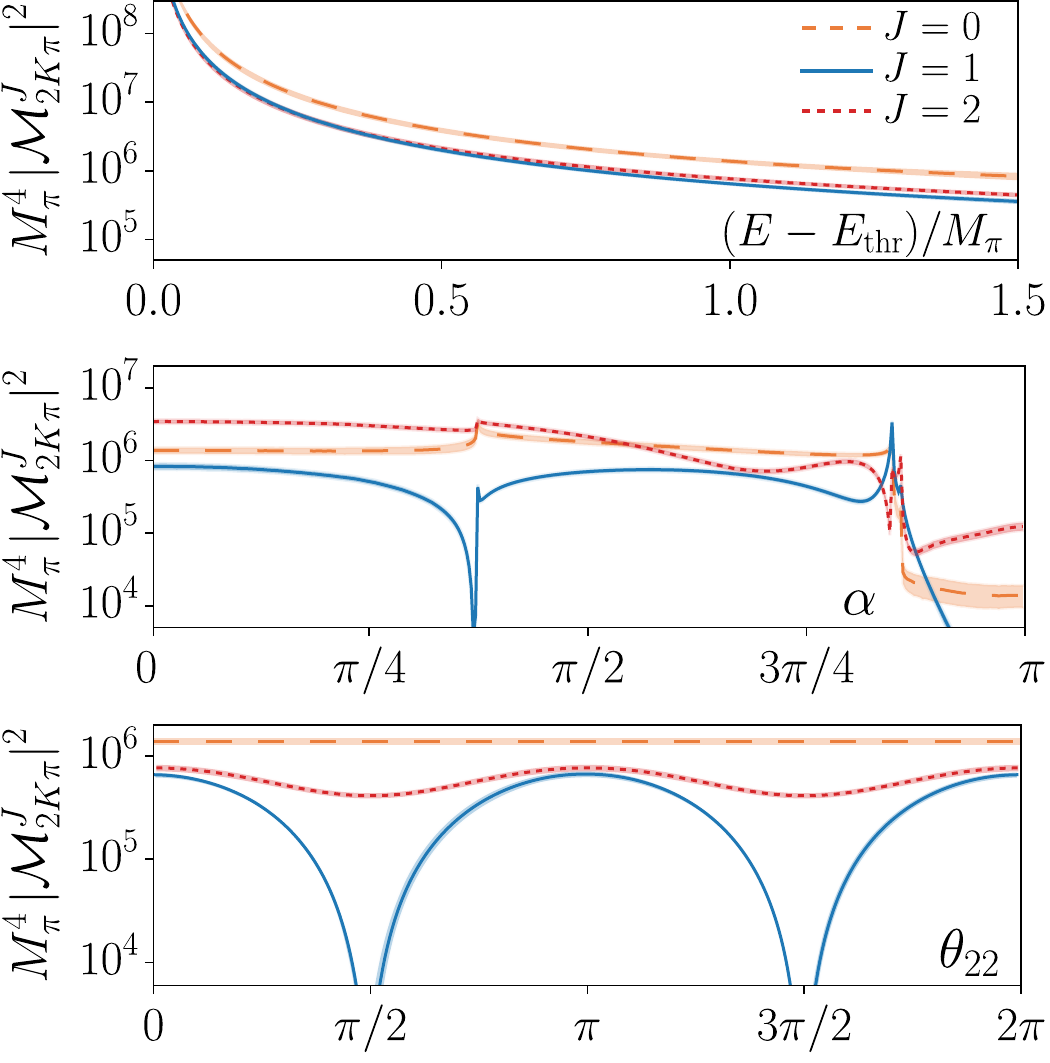}
    \vspace{-0.1cm}
    \caption{ 
    Squared magnitude of the $2K^+\pi^+$ amplitude for angular momentum $J=0,1,2$ as a function of three kinematic variables. From top to bottom: $E$ dependence for $\theta_{22} = 0$ and $\alpha = 120^\circ$; $\alpha$ dependence for $E = 2(M_K + M_\pi)$ and $\theta_{22} = 0$;  $\theta_{22}$ dependence for $E = 2(M_K + M_\pi)$ and $\alpha = 120^\circ$.
    } \vspace{-0.4cm}
    \label{fig:amps}
\end{figure}

\begin{figure*}[th!]
    \centering
    \includegraphics[width=\textwidth]{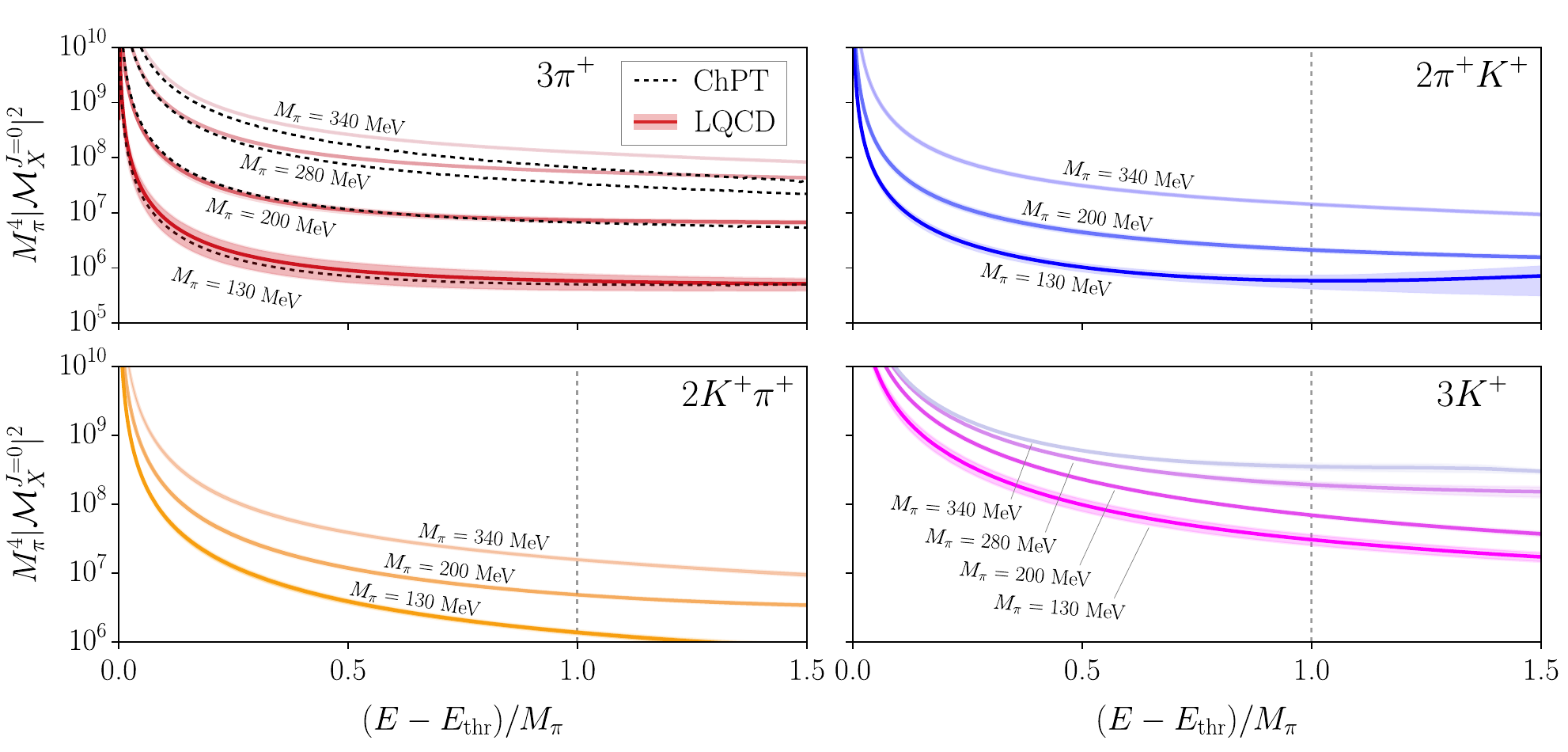} 
    \vspace{-0.44cm}
    \caption{ $J^P=0^-$ three-meson scattering amplitude squared as a function of the energy difference to threshold for different systems and ensembles. Incoming and outgoing momenta are both chosen to lie in an equilateral triangle configuration. Results are independent of $\theta_{22}$. NLO ChPT predictions are shown for the three-pion system. The dashed vertical grey lines indicate the lowest inelastic threshold. For $3\pi$, this is beyond the plotted region.
    } \vspace{-0.4cm}
    \label{fig:amps_J0_masses}
\end{figure*}

{\it Three-meson amplitudes.}---Having constrained $\cK_2$ and $\cK_3$ from LQCD energies, we solve the integral equations to obtain elastic three-meson scattering amplitudes. We decompose the three-meson amplitude in partial waves
\begin{align}
    \cM_3(\{\bm p\}; \{\bm k\}) = \sum_J \cM_3^J(\{\bm p\}; \{\bm k\}) \, ,  
    \label{eq:symmetrization}
\end{align}
where $\{\bm k\}$ and $\{\bm p\}$ represent the set of three initial and final momenta, respectively. We study fixed-$J$ amplitudes, $\cM_3^J$, which depend on eight independent variables after considering Poincar\'e invariance. To obtain a two-dimensional representation, we impose additional constraints by fixing certain relative orientations among the momenta---a process we refer to as ``choosing a kinematic configuration.''

We show our chosen kinematic configuration in \Cref{fig:config}. Both the incoming and outgoing momenta lie in the same plane. Each momentum lies along the bisector of a triangle: an equilateral triangle for the incoming momenta, and an isosceles triangle for the outgoing. The angle \(\theta_{22}\) specifies the relative orientation between the incoming and outgoing triangles, while the angle \(\alpha\) is measured between the two outgoing momenta of equal length (\(\alpha = 120^\circ\) recovers an outgoing equilateral triangle).

The panels of \Cref{fig:amps} shows the squared amplitude in three different partial waves for the $2K^+\pi^+$ system as a function of either $E$, $\alpha$, or $\theta_{22}$ while fixing the remaining two variables. Several features can be observed.
First, when considered as functions of the $E$, all amplitudes diverge at threshold due to one-particle exchange between two pairwise scatterings, an effect dominating all solutions of three-body integral equations. Second, the dependence of the definite-$J$ amplitudes on $\alpha$ reveals logarithmic singularities (rounded into cusps by numerical approximations) precisely at the transition point where on-shell one-particle exchanges become kinematically allowed. Third, the $\theta_{22}$ dependence  is flat for $J=0$ and exhibits the expected periodicity for $J=1$ and $2$. 

{\it Quark mass dependence.}---The quark mass dependence of the three-meson systems is investigated in \Cref{fig:amps_J0_masses}. Due to the normalization by the pion mass, all amplitudes become smaller as the pion mass is reduced. However, amplitudes involving pions contain an additional suppression, consistent with expectations from ChPT.

Given the lack of experimental results for such systems, the only source of guidance comes from effective theories, specifically ChPT. We plot the NLO ChPT result for the $3\pi^+$ system, obtained by applying the Inverse Amplitude Method~\cite{GomezNicola:2001as} and the three-body integral equations to the results of Refs.~\cite{Bijnens:2021hpq,Bijnens:2022zsq,Baeza-Ballesteros:2023ljl,Baeza-Ballesteros:2024mii}, and compare it to the LQCD results. As can be seen, better agreement is observed for lower pion masses and closer to threshold, which is consistent with the power counting of the chiral expansion. 

{\it Conclusions.}---We have demonstrated that LQCD calculations have reached a stage where physical-point three-particle scattering amplitudes can be determined, both in the lowest and higher waves.
LQCD thus provides a complementary tool in the study of resonances---accessing
direct three-to-three scattering processes, while experimental measurements are necessarily less direct. 

Given the success of three-hadron studies directly at the physical point, next steps will involve improved control of systematic uncertainties and studies of more complex systems with resonances or baryons.

\vspace{2mm} 
\noindent 
{\bf Acknowledgements}:
We acknowledge helpful discussions with Jorge Baeza-Ballesteros, Andrew Jackura, Maxim Mai, and Arkaitz Rodas.
We thank our colleagues within the CLS consortium for sharing ensembles,
and Marco C\`e for providing data for the meson decay constants on E250.

The work of SMD, ZTD, and SRS is partially supported by U.S. Department of Energy grant No. DE-SC0011637. FRL acknowledges partial support by the Mauricio and Carlota Botton Fellowship and by the USDOE Contract No. DE-SC0011090 and DE-SC0021006. CM and SS acknowledge support from the U.S. NSF under award PHY-2209167.

Quark propagators were computed on Frontera at the Texas Advanced Computing Center (U.S. NSF, award OAC-1818253) and Hawk at the High Performance Computing Center in Stuttgart. Meson functions and contractions were carried out on ``Mogon II" at JGU Mainz.

\bibliography{cited_refs}   

\end{document}